# Side Eye: Characterizing the Limits of POV Acoustic Eavesdropping from Smartphone Cameras with Rolling Shutters and Movable Lenses

Yan Long*, Pirouz Naghavi†, Blas Kojusner†, Kevin Butler†, Sara Rampazzi†, and Kevin Fu‡

*Electrical Engineering and Computer Science, University of Michigan
† Computer and Information Science and Engineering, University of Florida
‡ Electrical and Computer Engineering and Khoury College of Computer Sciences, Northeastern University
yanlong@umich.edu, {pnaghavi, bkojusner, butler, srampazzi}@ufl.edu, k.fu@northeastern.edu

*Abstract*—Our research discovers how the rolling shutter and movable lens structures widely found in smartphone cameras modulate structure-borne sounds onto camera images, creating a point-of-view (POV) optical-acoustic side channel for acoustic eavesdropping. The movement of smartphone camera hardware leaks acoustic information because images unwittingly modulate ambient sound as imperceptible distortions. Our experiments find that the side channel is further amplified by intrinsic behaviors of Complementary Metal-oxide–Semiconductor (CMOS) rolling shutters and movable lenses such as in Optical Image Stabilization (OIS) and Auto Focus (AF). Our paper characterizes the limits of acoustic information leakage caused by structure-borne sound that perturbs the POV of smartphone cameras. In contrast with traditional optical-acoustic eavesdropping on vibrating objects, this side channel requires no line of sight and no object within the camera's field of view (images of a ceiling suffice). Our experiments test the limits of this side channel with a novel signal processing pipeline that extracts and recognizes the leaked acoustic information. Our evaluation with 10 smartphones on a spoken digit dataset reports 80.66%, 91.28%, and 99.67% accuracies on recognizing 10 spoken digits, 20 speakers, and 2 genders respectively. We further systematically discuss the possible defense strategies and implementations. By modeling, measuring, and demonstrating the limits of acoustic eavesdropping from smartphone camera image streams, our contributions explain the physics-based causality and possible ways to reduce the threat on current and future devices.

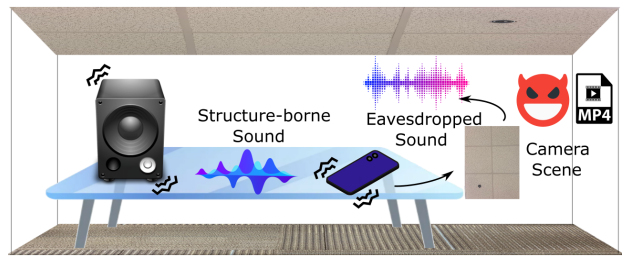

Fig. 1: Illustration of the POV optical-acoustic side channel when a camera is recording a ceiling or floor. Adversaries can eavesdrop structure-borne sounds emitted by electronic speakers by extracting acoustic signals from artifacts of lens movement and rolling shutter patterns in smartphone cameras that depend on POV rather than objects in the field of view.

## I. INTRODUCTION

Smartphone and Internet of Things (IoT) cameras are increasingly omnipresent near sensitive conversations even in private spaces. Our work introduces the problem of how to prevent the extraction of acoustic information that is unwittingly modulated onto image streams from smartphone cameras. We center our analysis on a discovered point-of-view (POV) optical-acoustic side channel that leverages unmodified smartphone camera hardware to recover acoustic information from compromised image streams. The side channel requires access to an image stream from a smartphone camera whose lens is near the eavesdropped acoustic source emitting structure-borne sound waves. The key technical challenge is how to characterize the limit of partial acoustic information leakage from humanly imperceptible image distortions, which is made possible by nearly universal movable lens hardware and CMOS rolling shutters that are sensitive to camera vibrations.

The most related body of research on optical-acoustic side channels involves recording videos of vibrating objects within the field of view with specialized, high-frame rate cameras [19], [32], [78], [79], [81]. However, innovations in privacy-aware camera systems and software can actively detect and hide sensitive objects in camera images to prevent such direct data leakage [31], [69], [74]. In contrast, our work explores the optical-acoustic side channel intrinsic to existing smartphone camera hardware itself, eliminating the need for objects in the field of view or line of sight: an image stream of a ceiling suffices (Figure 1). That is, we extract acoustic information from the vibratory behavior of the built-in camera—rather than the behavior of a vibrating object within the field of view of a specially mounted camera.

Our threat model and approach build upon previous research that used smartphone motion sensors for acoustic eavesdropping [20]–[22], [27], [46], [62], where structure-borne sound emitted by electronic speakers vibrates motion sensors and also leaks acoustic information. However, cameras do not directly encode acoustics like motion sensors. Instead, our work must demodulate acoustic information unwittingly encoded within image stream artifacts. Assessing the limits of information recovery with this optical-acoustic side channel thus poses the challenge of designing a signal processing pipeline that optimizes (1) the acoustic signal extraction from images and (2) the effective utilization of extracted signals. To tackle the first challenge, we characterize the side channel's signal path and model the rolling shutter pattern formation under sound wave motions as a signal modulation process. Our modeling reveals the limits of recoverable signal posed by factors such as imaging exposure time that can be optimized. It also reveals the theoretical signal extraction process, which guides us to design a diffusion registration-based extraction algorithm that rapidly and robustly recovers sound signals. Our recovered signals[2] with mainstream smartphones preserve over 600 Hz bandwidth of speech spectrum.

To tackle the second challenge, we observe that the extracted band-limited signals are complex transformations of the original sound and thus difficult for humans to recognize directly. In order to fully utilize the information embedded in the extracted signals, we design a classification model based on the HuBERT Large transformer [38]. Our extensive evaluation with 10 smartphones on a widely used spoken digit dataset [25] suggests that this optical-acoustic side channel powered by our signal processing pipeline allows adversaries to recover acoustic information from the surroundings. Specifically, we observed 80.66% accuracy on speaker-independent 10-digit recognition, 91.28% accuracy on recognizing 20 speakers, and 99.67% on gender recognition when a Google Pixel 3 phone was placed beside a speaker on a desk. In addition to classification, we also used NIST-SNR and Short-Time Objective Intelligibility (STOI) metrics to measure the quality and intelligibility of recovered speech signals, and observed scores up to 28 dB and 0.53, respectively. We further evaluated this side channel's robustness with various speaker volumes and speaker-phone distances as well as its applicability in different structure-borne propagation scenarios, including when the phone and speaker are placed on different desks or in different rooms.

Finally, we systematically investigate the possible defenses from the standpoints of user-based countermeasures and future camera design improvement respectively. For the latter, we propose corresponding hardware modifications to mitigate the two enabling factors of this attack, namely rolling shutter and movable lens. To summarize, the goal of this work is to model, measure, and demonstrate the capability of the POV optical-acoustic side channel on smartphone cameras and help defend against the threat on current and future camera devices. Our main contributions are summarized as follows:

- We identify and model smartphone camera characteristics that enable acoustic information to leak into camera

[2]Sample audio and additional materials can be found on our project website https://sideeyeattack.github.io/Website/

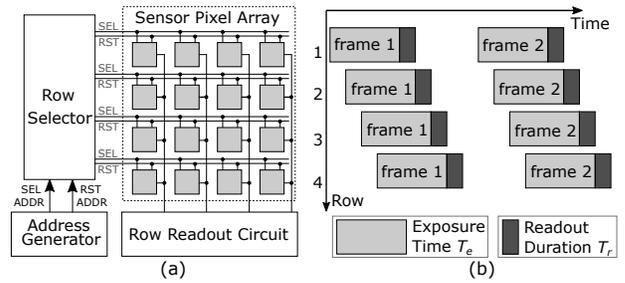

Fig. 2: (a) CMOS rolling shutter camera's row-wise sampling architecture with a $4 \times 4$ sensor pixel array. (b) The sequential readout of rows for two consecutive frames with exposure time $T_e$ and row readout duration $T_r$.

images through an optical-acoustic side channel. This POV optical-acoustic side channel unwittingly modulates ambient sound into image streams.
- We design a signal processing pipeline to characterize the limit of this side channel in several attack scenarios and evaluate it with 10 smartphones on a spoken digit dataset. It achieves 80.66%, 91.28%, and 99.67% accuracies on digit, speaker, and gender recognition respectively with a Google Pixel 3 placed beside a speaker on a desk. .
- We systematically analyze the possible defenses by investigating both user-based countermeasures such as adding dampening to phones, and camera design improvements that aim to address problems caused by rolling shutter and movable lens respectively. Our results indicate that mitigating lens movement is the most effective approach and combining multiple defenses can further reduce attack performance to the random-guess range.

## II. BACKGROUND

### A. Rolling Shutter Cameras

Rolling shutter cameras, which feature a *row-wise sampling architecture* (Figure 2 (a)), dominate the market of portable electronics including smartphones. Row-wise sampling overlaps the exposure of one row with the read-out of subsequent rows (Figure 2 (b)). An address generator controls this process by generating row-reset (RST) and row-select (SEL) signals that start the exposure and read-out of each row respectively. The interval between the two signals is the exposure time $T_e$. The duration of each row's read-out is denoted as $T_r$. The row-wise sampling architecture comes at the cost of additional image distortions when the optical paths change while imaging a scene. The optical paths can change when a relative movement happens between the scene, the lens, and the pixel array. The rolling shutter distortions are thus a function of optical path variations.

### B. Movable Lens

While the CMOS photo-sensitive pixel array is mounted on printed circuit boards (PCB) and rigidly connected to the camera body, the lens in most modern CMOS cameras is flexibly connected to the camera body by suspension structures using springs and specialized wires [60]. Such suspension structures allow relative movement between the lens and the pixel array, as shown in Figure 3. The movable lens is an essential component of cameras' optical image stabilization (OIS) and auto-focus (AF) systems and is almost ubiquitous in hand-held camera devices including smartphone cameras.

**Optical Image Stabilization:** OIS is an image stabilization method for mitigating tremor-caused motion blurs (Appendix A). Most OIS systems allow for 2D movements of the lens that are parallel to the pixel array plane, resulting in translational transformation of images. We only consider 2-DoF OIS movements and term such movements as XY-axes movements. OIS lens stroke is on the order of 100 μm [57].

**Auto-focus:** Most AF systems support 1-DoF movements of the lens on the axis that is perpendicular to the pixel array plane, which we term as Z-axis movements. Such movements can induce zooming effects that can be viewed as scaling transformations of the 2D image. AF lens stroke is also on the order of 100 μm [34].

**Sound Propagation.** This work investigates the consequences of movable lenses vibrated by structure-borne sound waves. Sound waves can propagate both through the air by inducing movements of air molecules, and through structures by inducing mechanical deformations in them. Structure-borne propagation can often transmit much higher sound energy than air-borne propagation [30]. In 2018, Anand et al. systematically analyzed the response of smartphone motion sensors to air-borne and structure-borne sound waves [20]. Their experiments show that structure-borne sound generated by electronic speakers causes stronger vibrations of the sensors and thus enables more feasible eavesdropping with motion sensors. Building upon their results, our work explores how structure-borne sounds can affect smartphone cameras.

## III. THREAT MODEL

### A. Problem Formulation

We characterize the threat of POV acoustic information leakage into smartphone cameras through structure-borne sound propagation. The sound generated by a sound source in the vicinity of a camera propagates to the camera and vibrates it, inducing rolling shutter effects in the camera image stream. The rolling shutter pattern thus becomes a function of the acoustic signal. The objective of an adversary is to learn the reverse mapping from the rolling shutter pattern to the privacy-sensitive information in the acoustic signal. Formally, we define the eavesdropping attack that an adversary $\mathcal{A}$ launches as a function $f_\mathcal{A}$:

$$f_\mathcal{A}: \quad \{P_v(S^l(t), \mathbb{E}), \mathbb{E}_\mathcal{A}\} \longrightarrow \tilde{l}, \quad \tilde{l} \in \mathbb{L}$$

where $S^l(t)$ is the continuous-time acoustic signal generated by the sound source; $l, \tilde{l} \in \mathbb{L}$ are the true and estimated information label of the acoustic signal; $\mathbb{L}$ is the set of all possible information labels and is reasonably assumed to be finite; $\mathbb{E}, \mathbb{E}_\mathcal{A}$ are the sets of environmental factors that are present during the attack (e.g., phone-speaker distance) and that are controlled or known by the adversary respectively, and have $\mathbb{E} \supseteq \mathbb{E}_\mathcal{A}$; $P_v(\cdot)$ denotes the projection from the acoustic signal to the videos containing the rolling shutter pattern. To measure the threat, we define the advantage of an adversary over random-guess adversaries as a probability margin

$$\mathbf{Adv}_\mathcal{A} = \mathbb{P}\left[f_\mathcal{A}(P_v(S^l(t), \mathbb{E}), \mathbb{E}_\mathcal{A}) - l < \epsilon\right] - \frac{1}{|\mathbb{L}|} \quad (1)$$

where $\epsilon$ is an arbitrarily small number. A successful attack is defined as $\mathbf{Adv}_\mathcal{A} > 0$. Although $\mathbf{Adv}_\mathcal{A}$ is a theoretical value that requires knowing the probability distributions and functions in Equation 1 to calculate, we can estimate this value by obtaining classification accuracies on datasets with equally likely labels as the ones in Section VI.

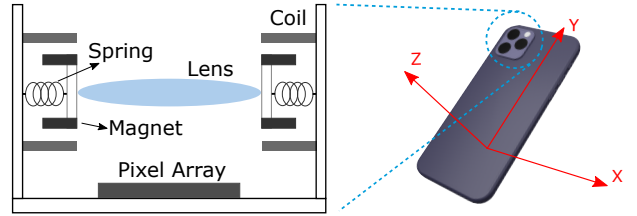

Fig. 3: The movable lens structure widely exists in smartphone cameras with optical image stabilization (OIS) and auto-focus (AF). When sound waves move the camera lens suspended on the springs, the optical path changes and creates an optical-acoustic side channel.

**Targeted Information Recovery.** We focus on recovering information from human speech signals broadcast by electronic speakers, as this is one of the most widely investigated threat models validated by previous research [20], [46]. In particular, our study investigates the feasibility and limit of recovering acoustic information from smartphone cameras without requiring microphone access. To better assess the limit, we allow the adversary to utilize state-of-the-art signal processing and machine learning techniques. We discuss three types of information recovery with increasing difficulty, namely (1) inferring the human speaker's gender, (2) inferring the speaker's identity, and (3) inferring the speech contents.

**Adversary Characteristics.** We consider an adversary in the form of a malicious app on the smartphone that has access to the camera but cannot access audio input from microphones (see Section III-B for the possible scenarios). In common mobile platforms including Android and iOS, the app will have full control over the camera imaging parameters such as focus and exposure controls once the camera access is granted. An adversary can change these parameters for optimal acoustic signal recovery based on their knowledge of the signal modulation process. We assume the adversary captures a video with the victim's camera while the acoustic signal is being broadcast. We also assume the adversary can acquire speech samples of the target human speakers beforehand to learn the reverse mapping to the targeted functions of the original speech signals and they can perform this learning process offline in a lab environment, which have been the standard assumptions in related side-channel research [21], [22], [46].

## B. Attack Scenario

Sounds broadcast by an electronic speaker can reach a smartphone's camera through structure-borne propagation when there exists a propagation path consisting of a single structure or a system of structures such as tables, floors, and even human body. Such a structure-borne model has been frequently used in previous works [20], [46], [72] of smartphone acoustic eavesdropping. Similar to previous works of motion sensor side channels, the malicious app eavesdrops on acoustic information under the general user expectation that no information can be stolen through sound when smartphone microphone access is disabled. Although camera access is usually regarded as being on the same privacy level as microphone access, users aware of the risk of acoustic leakage through microphones are still likely to grant camera access to apps until they realize the existence of the optical-acoustic side channel. We believe this can happen in three major situations. (1) The malicious app requests only camera access without microphone usage in the first place. Apps can disguise themselves as hardware information checking utilities (e.g., the widely used "AIDA64" app [5]) or silent video recording apps that do not record any audio. (2) The malicious app requests both camera and microphone access but a cautious user only grants camera access. We found that filming apps (e.g., the "Open Camera" [14] and "Mideo" [13]) often simply record without audio when microphone access is not granted. (3) The malicious app requests and is granted both camera and microphone access, but a user physically disables the microphone input by using external gadgets such as the Miclock microphone blocker [12]. Additionally, malicious apps can record videos stealthily without camera preview or in the background as has been done by existing apps like the "Background Video Recorder" on the Google Play Store [7] and "SP Camera" on the Apple App Store [16].

## IV. OPTICAL-ACOUSTIC SIDE CHANNEL

In this section, we seek to answer three key questions regarding the feasibility of the side channel: (1) Why does such a side channel happen? (2) What are the factors deciding the channel's capability? (3) How can adversaries extract high-quality signals from the channel?

### A. Signal Path Causality

**Mechanical Subpath.** When the electronic speaker on a table plays audio with total kinetic energy $E_s$, part of the kinetic energy it generates $k_0 E_s$ propagates to the body of the phone in the form of structure-borne sound waves and vibrates the smartphone body. Specifically, longitude waves mainly cause XY-axes motions of the smartphone body while transverse and bending waves mainly cause Z-axis motions [30]. The smartphone body and the camera body, including the sensor pixel array, are rigidly connected and thus have the same motion amplitude and velocity. Viewing them as a single unit separated from the camera lens, we denote the kinetic energy causing vibrations of this unit as $E_p$. We can approximately model this unit's motions on the table as a spring-mass system [68] with a spring constant $c_p$ and motion amplitude $A_p$. The camera lens is connected to the camera body through springs and can thus be regarded as a second spring-mass system. A portion of $E_p$, denoted as $k_1 E_p$, is converted to its elastic potential energy by stretching/compressing the springs. Denote the effective spring constant of the lens suspension system as $c_l$ and the relative motion amplitude between the lens and the smartphone-camera unit as $A_l$ ($A_l < A_p$), we then have

$$k_0 E_s = E_p = \frac{1}{2} c_p A_p^2 = \frac{1}{k_1} \frac{1}{2} c_l A_l^2 \qquad (2)$$

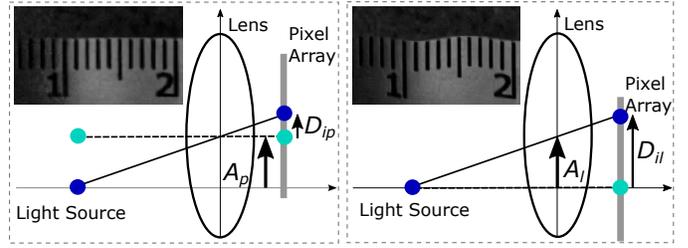

Fig. 4: The movable lens structure acts as a signal amplifier when structure-borne sound vibrates the smartphone camera. The dotted and solid lines represent the light ray projected before and after vibration. (Left) Without moving lenses, the rolling shutter pattern induces negligible pixel displacements. (Right) When lenses move, pixel displacements get amplified.

Note that $k_0$ and $k_1$ are frequency-dependent and reflect the physical properties of the mechanical system consisting of the speaker, the table, and the phone. In other words, $A_p$ and $A_l$ can be expanded along the frequency axis to represent the frequency response (transfer function) of the mechanical subpath. Such frequency response is hard to model but can be measured in an end-to-end manner (Section IV-D).

**Optical & Electronic Subpaths.** The movements of the smartphone body and the lens change the optical paths in different ways. Figure 4 shows a simplified model of how the two types of movements on the X-axis affect the light ray from a still point source to the sensor pixel array. In Figure 4 (a), the smartphone-camera body unit moves by $A_p$ while there exists no relative movement between it and the lens. With a focal length of $f$ (on the order of 5 mm[3]) and a camera-scene distance of $d$, the light ray projection point on the pixel array shifts by $\frac{f}{d} A_p$. In Figure 4 (b), only the lens is moving by $A_l$ while the smartphone-camera unit stays still. In this case, the projection point shifts by $(1 + \frac{f}{d}) A_l$. The optical projections are then sampled by the photo-sensitive pixel array and converted to digital signals, with the shifts of the projection point converted to pixel displacements in the images. Denote the general pixel displacement as $D_i$, the two types of movements will then result in pixel displacements of $D_{ip} = \frac{f}{d} \frac{A_p}{H} P$ and $D_{il} = (1 + \frac{f}{d}) \frac{A_l}{H} P$, where $H$ and $P$ are the

---

[3]The commonly claimed focal lengths on the order of 20 mm are the values converted to the equivalent of a full-frame camera sensor instead of the true physical values.

## B. Rolling Shutter Modulation

As pointed out in Section II-B, multi-DoF motions of the lens will mainly cause translation and scaling 2D transformations in the image domain. With a rolling shutter, transformations caused by multiple motions will be combined into one image frame because of the row-wise sampling scheme, and consequently produce wobble patterns that can be viewed as the outcome of modulating vibration signals onto the image rows. Furthermore, motion blurs exist due to the finite (namely, not infinitely small) exposure time of each row. For example, Figure 5 (b) and (c) show the simulated rolling shutter image ($250 \times 250$) with an exposure time of 1 ms when 500 Hz sinusoidal motion signals on the X and Z axis are modulated onto the image in Figure 5 (a) respectively. In light of these observations, we model the limits of acoustic signal recovery.

*1) Imaging Process:* We can model the imaging process of each row in a frame as a linear process where the final (row) image is the summation of different views that are 2D transformations of the original/initial view within the exposure time. The summation is actually the accumulation of photons on the CMOS imaging sensors. Consider frames of size $M$ rows, $N$ columns, and the simplest case where the motion only results in a uni-axis translation transformation on the column direction ($X$ axis). We denote the $i$-th row of the initial view as a column vector $r(i)$, and the matrix formed by all the possible translated views of $r(i)$ as $R_i = \begin{bmatrix} ... & r^{j-1}(i) & r^j(i) & r^{j+1}(i) & ... \end{bmatrix}$. Theoretically, $R_i$ has an infinite number of columns as the translation is spatially continuous. Considering a more practical discretized model, we let $j$ correspond to the displacement value in pixels in the image domain. For example, $r^{-3}(i)$ denotes the view shifted to the reverse direction along the $X$-axis by 3 pixels. Allowing negative indexing to $R_i$ for convenience and discretizing the continuous physical time with small steps of $\delta$, the formation of the $i$-th row in the $k$-th image frame, which is denoted as $\widetilde{r}(k, i)$, can then be expressed as the summation of different columns of $R_i$:

$$\begin{cases} \widetilde{r}(k,i) = \sum_{n=n_{k,i}^{start}}^{n \leq n_{k,i}^{end}} R_i[:, s(n\delta)] \\ n_{k,i}^{start} = \frac{T_f^k + (i-1)T_r}{\delta} \\ n_{k,i}^{end} = \frac{T_f^k + (i-1)T_r + T_e}{\delta} \end{cases} \quad (4)$$

where $T_f^k$ denotes the imaging start time of the frame and $s(n\delta)$ denotes the discrete motion signal with amplitude $D_i$ (Eq. 3) in the image domain. Equation 4 shows how rolling shutter exposure modulates the signal onto the images' rows. The objective of the adversary is to recover $s(n\delta)$ from $\widetilde{r}(k, i)$.

*2) Limits of Recovery:* With the modeling above, we can compute the characteristics of the recoverable signals.

**Captured Signal.** Signals in time intervals $[n_{k,M}^{end} \delta, n_{k+1,1}^{start} \delta]$, i.e., the gap between different frames, cannot be recovered since no camera exposure happens then. We term this portion

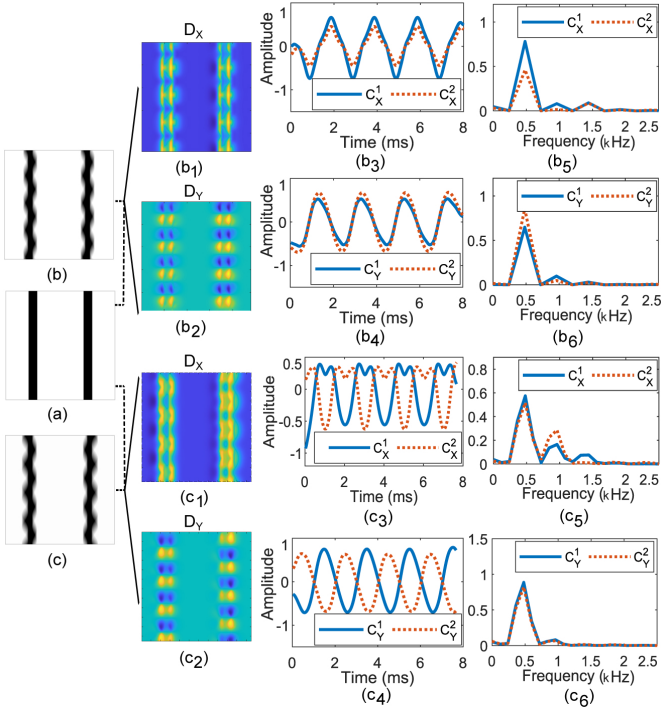

Fig. 5: The simulated rolling shutter images under a 500 Hz sound wave and the extracted signals with diffusion-based image registration. (a) The original scene. (b, c) The scenes with X and Z-axis motions respectively. (b/$c_{1,2}$) The X and Y-direction displacement fields. (b/$c_{3,4}$) The time domain signals computed from displacement fields with column-wise channels. (b/$c_{5,6}$) The corresponding frequency domain signals.

physical sizes and pixel resolution of the sensor pixel array on the X-axis respectively.

The interesting question arises as to whether $D_{ip}$ or $D_{il}$ is the main enabling factor of this side channel. Note that $\frac{f}{d}$ is very small since the camera-scene distance is usually larger than 10 cm. In light of this, we hypothesize $D_{il}$ is the dominant factor assuming $A_p$ and $A_l$, which cannot be measured directly, are on the same order of magnitude. We then verify our hypothesis experimentally by recording videos while preventing and allowing lens movements using a magnet. Figure 4 shows the significantly higher pixel displacement magnitudes when the lens is free to move under a 200 Hz sound wave. With a small distance $d$ of 10 cm, we observed $D_{ip} < 1px$ and $D_{il} \approx 8px$, which translates to $A_p < 63\,\mu m$ and $A_l \approx 22\,\mu m$. We thus ascertain that the lens movement is the main cause of the noticeable pixel displacements in the images. In other words, *the movable lenses act as motion signal amplifiers compared to those cameras that can only move with the smartphone body*. In light of this finding, we model the displacement as a function of the lens movement as

$$D_i \approx D_{il} = (1 + \frac{f}{d})\frac{A_l}{H}P \quad (3)$$

as the "lost signal" and the remaining portion as the "captured signal". We can calculate the percentage of the captured signal

$$\eta_{cap} = f_v M T_r \quad (5)$$

where $f_v$ is the video frame rate. Higher $\eta_{cap}$ means the adversary can recover more information from images.

**Sample Rate & Bandwidth.** For the captured signal, although the adversary wants to infer all the transformed views and thus recover all signals in time intervals $[n_{k,i}^{start}\delta, n_{k,i}^{end}\delta]$, it is impossible to know the order of these views' appearance because the photons from all the views are summed in the exposure time and the loss of order information is irreversible. Without the order information, the adversary can only reformulate Equation 4 as

$$\begin{cases} \tilde{r}(k,i) = R_i x(i) \\ x(i)_j = \sum_{n=n_{k,i}^{start}}^{n \le n_{k,i}^{end}} \mathcal{I}\{s(n\delta) == j\} \end{cases} \quad (6)$$

where $x(i)$ is a coefficient column vector whose $j$-th entry $x(i)_j$ represents how many times the translated view $r^j(i)$ appeared within the exposure time; $\mathcal{I}\{\cdot\}$ is the indicator function. Theoretically, with the measurable final image $\tilde{r}(k,i)$ and the matrix $R_i$ that can be approximately constructed using a still frame, $x(i)$ can be computed by solving the linear system in Equation 6. To recover a 1D motion signal that is a function of $s(n\delta)$, the adversary can estimate a synthetic motion data point $a(i)$ from $x(i)$ by taking the weighted average of $j$ with respect to $x(i)$:

$$a(i) = \frac{\sum_j j \times x(i)_j}{\sum_j x(i)_j} = \frac{1}{T_e/\delta} \sum_{n=n_{k,i}^{start}}^{n \le n_{k,i}^{end}} s(n\delta) \quad (7)$$

The adversary-measurable signal $a(i)$ thus embeds the information of the original motion signal.

Based on Equations 4 and 7, we can conclude that the measurable signals extracted from the rolling shutter patterns have an effective sample rate of $1/T_r$. Equation 7 also shows that the sampling process from a motion-blurred image acts as a moving mean filter whose frequency response is determined by the exposure time $T_e$.

### C. Motion Extraction Algorithm

Directly using Equation 7 for signal extraction faces three real-world challenges: (1) Solving the linear system of Equation 6 is computation-intensive. (2) The size of $R_i$ increases exponentially as the motion's DoF increases. (3) Equation 6 is mostly underdetermined. We thus designed a motion signal extraction algorithm based on *diffusion-based image registration* [65], [70]. It takes in a reference image $I_{ref}$ and a moving image $I_{mov}$ of size $M \times N$ (number of rows and columns, e.g., $1080 \times 1920$), and outputs 2D displacement fields (matrices) for X and Y-direction displacements respectively, i.e., $D_X^{M \times N}$ and $D_Y^{M \times N}$. Figure 5 shows the raw displacement fields for (b) and (c). We further apply column-wise averaging to the matrices to reduce data dimensionality as well as the impact of random noise in the imaging process, which improves signal robustness. We assign columns to different groups and take group-wise averages on the X and Y displacement fields respectively. We empirically choose the number of groups $n_g$ to be the nearest integer to $2N/M$ to balance the robustness and the details we want to preserve. After averaging, we reduce $D_X$ and $D_Y$ to $4N/M$ 1D signals of length $M$ (number of rows), and we term each 1D signal as a channel. Let $dir \in \{X, Y\}$ and $a^i$ denote the averaging column vector with its $j$-th entry denoted as $a_j^i$, the channels are then formally defined as

$$\begin{cases} C_{dir}^i = D_{dir} \cdot a^i, \quad i = 1, 2, \ldots, n_g \\ a_j^i = \frac{N}{n_g} \mathcal{I}\left\{\frac{n_g}{N}(i-1) < j \le \frac{n_g}{N} i\right\} \end{cases}$$

For the $250 \times 250$ images in Figure 5, the 4 channels and their spectrums are shown in ($b_{3-6}$) and ($c_{3-6}$). When dealing with a video, i.e., a sequence of images, we use the first frame as $I_{ref}$, and concatenate consecutive frames' channels to get the channel signals of the video. We use the same notation $C_{dir}^i$ to denote a video's channels.

### D. Feasibility & Attack Characterization

**Camera Scene.** Most smartphones have both front and rear cameras. Although some smartphone manufacturers such as Vivo have started to equip their front cameras with OIS [4], we focus on rear cameras in this work since more of them are equipped with OIS and AF. The rear camera has a certain scene while imaging. The scene can affect information recovery because their structures, textures, and distance from the camera can modify the characteristics of the light rays entering the camera. The scene changes with the smartphone's placement and location. As depicted in Figure 1, a phone on a table with an upward-facing rear camera often records a scene of the ceiling ("Ceiling Scene"); a downward-facing camera on a non-opaque surface such as a glass table often records a scene of the floor ("Floor Scene"). For simplicity we assume there are no moving objects in the scene.

For our preliminary analysis, we use a test setup with a KRK Rokit 4 speaker and a Google Pixel 2 phone held by a flexible glass platform on a table with the phone's rear camera facing downwards to simulate a Floor Scene. We use a customized video recording app that acts as the malicious app (see Appendix A for detail) to record in MP4 format. We first discuss the choice of adversary-controllable camera parameters and then discuss the environmental factors in order to characterize the envelope of the adversary's capability.

**Camera Control Parameters.** The frequency response of the side channel is determined by both the mechanical subpath and the camera control parameters of the malicious app that can be optimized by the adversary. We estimate the frequency response by conducting a frequency sweep test where we play the audio of a chirp from 50 to 650 Hz. We then aim to find the optimum response for our Google Pixel 2. Figure 9 (a) shows the best response where the maximum recovered chirp frequency is about 600 Hz. Specifically, we optimize the control parameters in the following ways whose detailed reasoning and implementation can be found in Appendix A: (1) Disable auto-exposure and reduce the exposure time (Section IV-B). (2) Disable optical and electronic image stabilization (OIS and EIS) and auto-focus (AF). (3) Minimize video codec compression. (4) Maximize pixel resolution. (5) Choose appropriate frame rates for each phone. Figure 9 (b-f) also show the responses when optimum settings are not achieved.

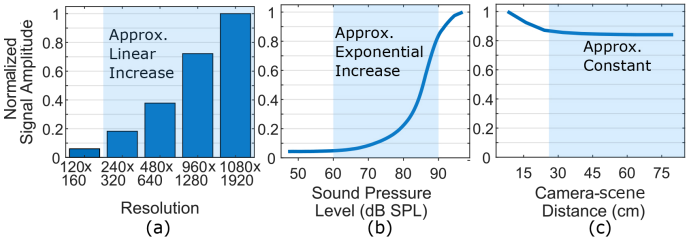

Fig. 6: The relationship between signal amplitudes (normalized) and different factors. (a) Amplitude increases approximately linearly with video resolution. (b) Amplitude increases approximately exponentially with speaker volume. (c) Amplitude remains approximately constant as the camera-scene distance changes due to the movable lens structure.

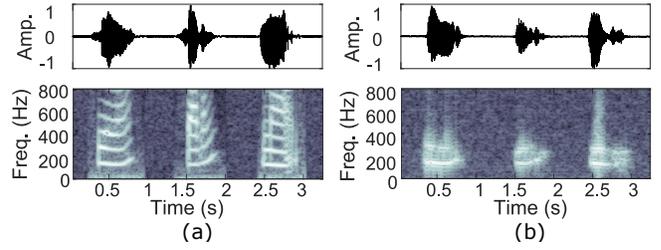

Fig. 7: The waveform and spectrogram of spoken digits "zero", "seven", and "nine". (a) The original signals. (b) The recovered signals from a 3.2s video with optimized camera parameters.

**Configuration Factors.** Variations of configuration factors can also affect the recoverable signals. We discuss the impact of three main factors: sound pressure, distance from the scene, and phone orientation.

(1) Sound pressure level. Louder sounds induce larger signal amplitudes, i.e., $D_i$ in Equation 3, by increasing $E_p$ and thus $A_p$. Figure 6 (b) shows that discernible signals appear when the SPL is larger than 60 dB. The signal amplitude increases exponentially as the SPL increases until the lens motion approaches the stroke limit of the suspension system around 90 dB. Such an exponential relationship agrees with our modeling in Section IV-A since the SPL is a logarithmic function of $E_p$ and $D_i \propto \sqrt{E_p}$. It suggests the attack might be relatively sensitive to changes in volumes. We will conduct further evaluations in Section VI-B.

(2) Camera-scene distance. According to Equation 3, the camera-scene distance $d$ has progressively smaller impacts on $D_i$ as it increases. Figure 6 (c) shows that the signal amplitude is approximately constant when the distance between the smartphone camera and the object in the scene is larger than 30 cm. Considering that both Ceiling and Floor Scenes often have distances much larger than 30 cm, Figure 6 (c) suggests this factor has a relatively small impact on the attack capability.

(3) Phone orientation. The orientation (on the XY-plane) of the phone with respect to the sound source changes the lens motion's directionality. We empirically evaluate how orientation can affect the attack by testing different orientations. We find that phone orientation has a relatively small impact on the extractable acoustic information since most cameras' movable lenses have at least 3 DoF. The lens motions in all directions can thus be effectively captured.

(4) Other factors. Besides the three factors above, other factors such as the phone-speaker distance affect the recovered signal in less quantifiable ways due to the lack of descriptive mathematical models. We will evaluate the impact of these factors in typical settings in Section VI-B.

## V. LEARNING THE FUNCTIONS OF SPEECH

Figure 7 (a) and (b) show the original and recovered speech signals of a human speaking "one", "seven", and "nine". While we could detect clear tones and their dynamics with more than doubled recoverable frequency range compared to that of smartphone motion sensor side channels reported so far (about 250 Hz maximum) [22], the recovered speech audio is still challenging for humans to recognize directly. We believe the reason is that the maximum bandwidth of 600 Hz often only captures the fundamental frequency ($F_0$) of vowels and voiced consonants while losing the second and third formants ($F_0, F_1$); signals from unvoiced consonants (>2 kHz) such as "f", "s", "k" will also be completely missing [26], [75]. It has been shown that an audio channel with a 1kHz bandwidth could only allow single-word recognition rates of less than 20% by humans [52]. Furthermore, Figure 9 shows certain low frequencies such as 200 Hz can generate higher frequency distortions that can contaminate the true high-frequency signals. This suggests a human hearing system-based attack $f_A$ is not likely to succeed. We also found existing machine-based Speech-to-Text engines such as Google Cloud [9], IBM Watson [10], and Apple voice assistant [11] unable to detect speech in the recovered signals. The observation motivated us to construct a more specialized $f_A$ for estimating the information recovery limits.

### A. Signal Processing Pipeline

As shown in Figure 8, our $f_A$ is a signal processing pipeline that consists of the following three stages.

(1) Signal extraction. The stage implements the extraction algorithm shown in Section IV-C. It accepts videos collected by the malicious app and outputs $2n_g$ (8 in the case of 1080p videos) channels of 1D signals.

(2) Pre-processing. It performs noise reduction, liveness detection, trimming, low-pass filtering, and normalization of the channels. As shown in Figure 9, the extracted signals contain non-trivial but spectral-static noise caused by different imaging and image registration noise. We thus first apply a background noise reduction step using the spectral-subtraction noise removal method in [53], [54]. We then conduct a channel-wise amplitude-based liveness detection that determines the start and end index of the contained speech signals. Afterward, we average the start and end indices of the channels and trim them to remove the parts without speech signals. We further apply a digital low pass filter with a cutoff frequency of 4kHz to get rid of the remaining high-frequency disturbances caused by camera imaging noise. Finally, we normalize the channels and pass them to the next stage.

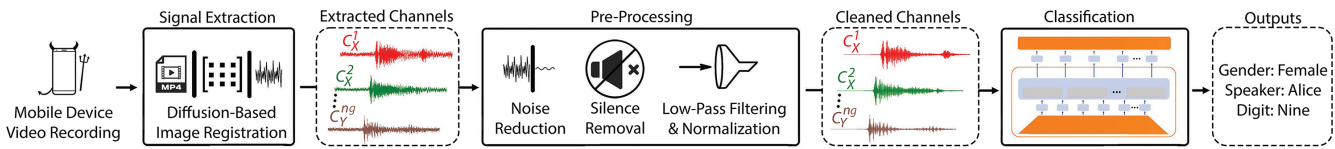

Fig. 8: Our signal processing pipeline exploits the optical-acoustic side channel on smartphone cameras. The signal extraction stage extracts sound-induced signals from the videos recorded on smartphones. The pre-processing stage cleans up the signals and feeds them into the classification model, where the gender, speaker, and speech content are recognized.

(3) Classification. Our classification stage implements a classification model that builds upon the Hidden-unit Bidirectional Encoder Representations from Transformers (HuBERT) large [38], which is introduced next.

### B. Classification Model

Our HuBERT-based classification model utilizes the advantages of transfer learning, waveform input, and state-of-the-art performance[2]. The model consists of three major components: CNN encoder, transformer, and classifier. To adopt the original HuBERT for our purpose, we change the model by (1) modifying the CNN encoder to allow multiple waveform channels, (2) changing the transformer dropout probability, and (3) adding a classification layer to allow HuBERT to be used for spoken digit classification. We implement all of these changes while preserving as much of HuBERT's pre-training as possible to leverage the benefit of transfer learning. Preserving the pre-trained weights is particularly important for the CNN encoder because it helps avoid the vanishing gradient problem that commonly occurs when training deep neural networks [66]. We use the weights of the first layer for each channel of our input signal $C_X^1, ..., C_Y^{n_g}$ and change the original dropout probability of 0.1 to 0.05 to better regularize the model for our task. We then designed and added our classifier to process the output of the transformer. The classifier averages the non-masked discovered hidden units and outputs label scores for each classification task. In our classification tasks, gender, digit, and speaker output 1, 10, and 20 scores respectively, which are used to obtain the likelihood of each label and thus the final predicted class.

The CNN encoder contains 7 CNN layers, each outputting 512 channels. The first CNN layer inputs a single channel while the remaining layers input 512 channels. Since our input signal, $C_X^1, ...C_X^{n_g}, C_Y^1, ..., C_Y^{n_g}$, consists of multiple waveforms, the CNN encoder is modified accordingly, using all the input channels to discover utterances $Y_1, Y_2, ..., Y_m$. The transformer contains 24 blocks, an embedding size of 1024, and 16 attention heads, which amounts to 317 million trainable parameters. The generated hidden units output by the model can be used for a variety of speech recognition tasks. In the case of classification, the final classifier layer averages the non-masked discovered hidden units $Z_1, Z_2, ..., Z_m$ and outputs label scores for each classification task. We use cross entropy as the objective function for our binary and multi-class classification tasks during training. In hyperparameter tuning, we discovered that the initial learning rate of 1e-4 and a scheduler decaying it every 4 epochs by a factor of 0.8 delivers optimal results.

### VI. EVALUATION

To gauge the general capability of the optical-acoustic side channel, we carry out evaluations on a spoken digit dataset used in previous work of smartphone motion sensors acoustic side channel [22]. We first evaluate the structure-borne side channel's performance in shared-surface and different-surface scenarios separately using a Google Pixel 2 to investigate the impact of different structures and structure organizations, and then compare the performance between different phone models. For evaluation metrics, we provide both common speech audio quality metrics including NIST-SNR and Short-Time Objective Intelligibility (STOI), and accuracies of our specialized classification model. The former measures how good the extracted audio signals are and are used in major previous works of acoustic recovery & eavesdropping [32], [43], [47]–[49]. The latter measures how well information labels are extracted from audio signals to quantify the limits of information recovery. We found the two systems of metrics generally agree with each other as we observed correlation scores of 0.72 and 0.80 between our model's digit classification accuracies and NIST-SNR and STOI respectively with our evaluation data.

### A. Evaluation Setup

**Dataset & Classification Tasks.** The dataset is a subset of the AudioMNIST dataset[4] [25] and contains 10,000 samples of signal-digit utterances (digit 0-9) from 10 males and 10 females. We perform three classification tasks, namely speaker gender recognition, speaker identity recognition, and speaker-independent digit recognition. These three tasks correspond to the three levels of information recovery in Section III with $|\mathbb{L}| = 2, 20, 10$ respectively. Since all data labels for each task are equally likely in the dataset, the classification accuracies then serve as a statistical indication of $\mathbf{Adv}_\mathcal{A}$.

---
[4] https://github.com/soerenab/AudioMNIST

**Experimental Setup & Data Collection.** As our baseline setup, we place the smartphones and a KRK Classic 5 speaker side by side on a glass desk (Floor Scene), as shown in Figure 11 (a). The speaker volume measures 85 dB SPL at 1 m [63]. The impact of smaller volumes including normal conversation volumes will be discussed in Section VI-B. For each evaluation case, we collect the whole 10k samples in our dataset using Python automation. We randomize the order of collected samples to avoid biased results due to unknown temporal factors. All phones use an exposure time of 1 ms.

**Training & Classification Metric.** To train the HuBERT large model for classification, we randomly split the 10k-sample dataset into training, validation, and test sets with 70%, 15%, and 15% splits, respectively. For each device or scenario evaluation, we train 3 HuBERT large models, one for each classification task. We trained all the models from the original pre-trained HuBERT large to allow for better comparison and used the same test set for final evaluations of all the models, where we report classification accuracies on the test set. The validation set is used for hyperparameter tuning and final model selection. During training, the model with the highest Receiver Operating Characteristic Area Under Curve (ROC-AUC) score is selected as the final model.

**NIST-SNR and STOI.** NIST-SNR [2] (referred to as SNR hereafter) measures the speech-to-noise ratio by calculating the logarithmic ratio between the estimated speech signal power and the noise power. A higher SNR score indicates better signal quality. STOI [64] is a widely used intelligibility metric. A higher STOI score indicates the speech audio is more comprehensible to humans. For all evaluation cases, we measure the SNR and STOI over the 1536-sample test set to make it comparable to the classification accuracies reported. We also utilize SNR and STOI to measure signal quality in certain test cases that do not present a unique evaluation dimension by using a 100-sample signal testing subset consisting of 100 speech samples randomly sampled from the test set. In comparison to a full evaluation case, using the signal testing subset allows us to assess signal quality in a large number of different test cases in an efficient way. According to the sample size selection guideline from NIST [1], a sample size of 100 allows us to estimate the change in the average SNR and STOI scores with a 99% confidence level at a resolution of 0.5 times the standard deviation of the test set population's scores. With all the evaluation data we collected, this gives us a resolution of about 1.6 for SNR and 0.1 for STOI.

### B. Shared-surface Scenarios

Shared-surface scenarios include the phones and speakers on the same surface, usually a table. In different scenarios, the quality of the recovered signals varies with configuration changes as shown in Section IV-D. We first study the impact of camera scenes and speaker volumes individually, and then investigate several representative scenarios that incorporate different combinations of the key factors of surface structure and phone-speaker distance.

TABLE I: Performance In Shared-surface Scenarios

| Scenario | Case | Avg. SNR | Avg. STOI | G (%) | S (%) | D (%) |
|---|---|---|---|---|---|---|
| **Scene** | Floor Scene 1 | 18 | 0.51 | 99.87 | 91.02 | 79.69 |
| | Floor Scene 2 | 13 | 0.48 | 99.54 | 83.85 | 70.05 |
| | Ceiling Scene | 9 | 0.38 | 99.87 | 86.27 | 67.64 |
| **Volume** | 85 dB | 18 | 0.51 | 99.87 | 91.02 | 79.69 |
| | 75 dB | 11 | 0.44 | 99.80 | 89.13 | 76.95 |
| | 65 dB | 4 | 0.18 | 98.83 | 76.11 | 68.16 |
| | 55 dB | 2.4 | 0.13 | 80.27 | 34.77 | 27.67 |
| | 45 dB | 2.3 | 0.15 | 54.49 | 8.92 | 13.28 |
| | 35 dB | 2.3 | 0.14 | 54.23 | 6.84 | 15.95 |
| **Glass Desk, Distance, Volume** | 10 cm, 85 dB | 9 | 0.38 | 99.87 | 86.27 | 67.64 |
| | 10 cm, 65 dB | 1.9 | 0.25 | 81.25 | 37.17 | 32.03 |
| | 110 cm, 85 dB | 9.3 | 0.35 | 99.74 | 84.24 | 64.13 |
| | 110 cm, 65 dB | 1.8 | 0.32 | 81.12 | 36 | 31.12 |
| **Wooden Desk, Distance, Volume** | 10 cm, 85 dB | 4.4 | 0.19 | 98.37 | 73.11 | 57.55 |
| | 10 cm, 65 dB | 1.8 | 0.25 | 60.29 | 13.22 | 17.25 |
| | 130 cm, 85 dB | 5.2 | 0.22 | 99.48 | 83.59 | 69.53 |
| | 130 cm, 65 dB | 1.8 | 0.21 | 75.2 | 30.08 | 28.26 |
| **Wooden CR TBL, Distance, Volume** | 10 cm, 85 dB | 8.8 | 0.33 | 99.02 | 79.82 | 66.6 |
| | 10 cm, 65 dB | 2.4 | 0.19 | 76.76 | 42.58 | 32.49 |
| | 200 cm, 65 dB | 2.3 | 0.19 | 70.75 | 33.53 | 26.43 |
| | 300 cm, 65 dB | 2.6 | 0.19 | 83.2 | 41.86 | 30.99 |

TBL - Table, CR - Conference room, G - Gender, S - Speaker, D - Digit

**Camera Scene.** Table I shows the classification results under three scenes as shown in Figure 10. The first scene (Floor Scene 1) is with a downward-facing camera on the glass desk imaging the floor covered by a carpet. The second scene (Floor Scene 2) uses the same table and downward-facing camera but contains a different carpet on the floor. The third scene (Ceiling Scene) is with the camera upward-facing on the same table imaging the ceiling. Floor Scene 1 produces the highest accuracies in all three classification tasks, which we believe is due to the following reasons. First, the carpet in Floor Scene 1 has a lighter color than the carpet in Floor Scene 2, enabling more photons to be reflected and enter the camera and thus increasing the signal-to-noise ratio. Second, the image scene of Floor Scene 1 has larger contrast than that of the Ceiling Scene due to the more abundant textures of the carpet compared to the ceiling.

**Volume.** Different speaker volumes represent different daily scenarios. Figure 6 (b) shows that the speaker volume has a strong impact on the signal amplitude. We found, however, the sharp decrease in signal amplitude does not lead to a proportional decrease in the classification accuracies. Table I shows the result with 4 typical conversation volumes and 2 whisper/background volumes: 85, 75, 65, 55, 45, and 35 dB often represent shouting, loud conversation, normal conversation, quiet conversation, whisper, and background noise respectively [8]. The results indicate that for volumes of 55 dB and above, the $f_\mathcal{A}$ designed still has a significant advantage over a random-guess adversary, demonstrating the side channel's effectiveness in quiet conversation volumes. The accuracies appear to be in the random-guess range at 45 and 35 dB.

TABLE II: Performance With Different Speaker Devices

| Speaker Device | Avg. SNR | Avg. STOI | Gender (%) | Speaker (%) | Digit (%) |
|---|---|---|---|---|---|
| KRK Classic 5 | 18 | 0.51 | 99.87 | 91.02 | 79.69 |
| Logitech Z213 | 18 | 0.44 | 99.09 | 88.8 | 77.67 |
| Laptop G9-593 | 3.3 | 0.12 | 94.92 | 57.03 | 36.78 |
| Samsung S20+ | 6.4 | 0.15 | 89.00 | 53.91 | 32.36 |

TABLE III: Performance In Different-surface Scenarios

| Scenario | Avg. SNR | Avg. STOI | G (%) | S (%) | D (%) |
|---|---|---|---|---|---|
| Monitor Stand 85 dB | 11 | 0.45 | 99.09 | 80.53 | 60.42 |
| Monitor Stand 65 dB | 2.6 | 0.09 | 84.05 | 42.32 | 32.1 |
| Two Desks 85 dB | 2.6 | 0.08 | 75.72 | 19.6 | 14.26 |
| Two Rooms 85 dB | 2.3 | 0.06 | 66.93 | 15.17 | 15.17 |
| Shirt Pocket 85 dB | 2.5 | 0.19 | 95.9 | 66.37 | 45.7 |
| Bag Pocket 85 dB | 4.1 | 0.23 | 93.1 | 40.1 | 55.34 |

G - Gender, S - Speaker, D - Digit

**Surfaces Structure and Phone-speaker Distance.** Besides the glass desk, we evaluated a wooden desk in the same office and a 3m-long wooden conference room table. The Ceiling Scene was used for this set of evaluations. Table I shows the results with two different distances on the wooden and glass desks at 85 and 65 dB. The first distance is 10 cm and represents the scenario of placing the phone right beside the speaker; the other distance is the maximum achievable distance on each table (110 and 130 cm) by placing the phone on one edge and the speaker on the other edge, as shown in Figure 11. With the glass desk, a 3% decrease was observed for digit recognition when the distance increases from 10 cm to 110 cm. For the wooden table, the accuracies increased when the distance increased from 10 cm to 130 cm. Although this may seem counterintuitive at first, a closer look at the desks' mechanical structures suggests it is due to the smaller effective thickness on the edge of the table (Appendix B). At 65 dB, the glass and wooden desks show larger drops in accuracies than those in the volume experiments, which we believe is due to the ceiling scene having a more uniform color spectrum compared to Floor Scene 1, making smaller vibration amplitudes a more significant factor on classifier performance.

To further evaluate the side channel's robustness with larger phone-speaker distances, we conducted experiments with a 3m-long wooden conference room table. As shown in Table I, the classifiers' accuracies remain larger than random-guess accuracies, indicating the side channel's effectiveness at distances larger than 100 cm at normal conversation volumes.

**Speaker Device.** To uncover the potential impact of speaker devices on the side channel, we tested 4 different speaker devices including two standalone speakers (KRK Classic 5 and Logitech Z213), a laptop speaker (Acer Laptop G9-593), and a smartphone speaker (Samsung S20+). Table II shows that all 4 speaker devices allow for performance better than a random-guess adversary. We found even smaller internal speakers of portable devices including the laptop speaker vibrating a nearby phone's camera and the Samsung S20+'s speaker vibrating its own onboard camera could induce discernible signals. The variation in accuracies over the 4 devices is mainly due to the different maximum output volumes they can achieve; while the KRK Classic 5 and the Logitech Z213 speakers can output 85 and 75 dB respectively, the Laptop G9-593 and Samsung S20+ speakers are limited to 60 dB output.

**Additional Objects on Surface.** Thus far, most experiments were conducted with the speaker and the phone as the only objects present on the surface. Theoretically, the presence of additional objects on the surfaces propagating sound waves will only have a small impact on the side channel because structure-borne sound vibrates the entire structures which are often much heavier than the objects on the surfaces. To further investigate this factor, we conducted experiments with a daily occurring scenario of a cluttered desk with a varying set of common objects placed on the desk including a speaker, a laptop, a monitor, and a printer (Figure 12). Despite the slight change in SNR and STOI scores (Table VI), full evaluations of the least and most cluttered scenarios reported similar classification accuracies: the least cluttered scenario achieved 94.86%, 70.44%, and 50.98% for gender, speaker, and digit classification accuracy respectively while the most cluttered desk scenario achieved 91.41%, 69.27%, and 56.25%. The results suggest cluttered surfaces with heavy objects allow for similar side channel performance.

### C. Different-surface Scenarios

We evaluated several different-surface scenarios (Figure 13) including (1) the speaker on the desk and the phone on the desk's monitor stand; (2) the speaker on the floor and the phone in the pocket of a shirt and a backpack worn by a mannequin; (3) the speaker and phone on different desks; (4) the speaker and phone in separate rooms. Table VII indicates the side channel's performance over a random-guess adversary in these scenarios. With the exception of monitor stand experiments, we believe the decrease in performance can be attributed to the fact that the same speaker energy $E_s$ now vibrates structures of much larger weight and stiffness (in this case the concrete floor) as opposed to a wooden floor structure [77] or wooden/glass surface (Appendix B). This makes it more difficult to create oscillation of structures with larger amplitudes to produce higher SNR. Additional causes of performance degradation could be due to the contact point between the desk, the mannequin's foot, and the transfer medium, i.e., the floor, moving relative to each other and causing frictional losses of the vibration energy $E_s$ and thus also result in a lower SNR.

### D. Different Smartphones

To evaluate the capability and robustness of our side channel on different phones, we analyzed the classification accuracies of 10 phones in the Floor Scene 1 setup. Table IV shows the results from three smartphone families, namely the Google Pixel, Samsung Galaxy, and Apple iPhone. We also show the estimated recoverable frequency ranges, the rear camera modules, and their key characteristics in Table IX. To measure the key characteristics, we generate a 200 Hz tone for 3 seconds. We then find $1/T_r$ by changing it to align the recovered signal with 200 Hz. With $1/T_r$, we calculate $\eta_{cap}$ according to Equation 5. We further measure $\eta_{cap}$ by dividing the length of the recovered tone by 3 seconds. The measured and calculated $\eta_{cap}$ match well with each other which shows the correctness of our modeling. We used 30 fps for the Android phones because that is what most Android manufacturers currently provide to 3rd party apps while iPhone used 60 fps.

TABLE IV: Performance With Different Smartphone Models

| Device | Avg. SNR | Avg. STOI | Gender (%) | Speaker (%) | Digit (%) |
|---|---|---|---|---|---|
| Pixel 1 | 18 | 0.46 | 99.61 | 81.84 | 69.53 |
| Pixel 2 | 18 | 0.51 | 99.87 | 91.02 | 79.69 |
| Pixel 3 | 17 | 0.49 | 99.67 | 91.28 | 80.66 |
| Pixel 5 | 22 | 0.49 | 99.48 | 84.51 | 70.25 |
| Samsung S7 | 21 | 0.49 | 99.54 | 82.94 | 66.08 |
| Samsung S8+ | 17 | 0.45 | 99.61 | 79.30 | 57.29 |
| Samsung S20+ | 22 | 0.49 | 99.80 | 83.92 | 61.07 |
| iPhone 7 | 28 | 0.53 | 99.87 | 85.09 | 65.23 |
| iPhone 8+ | 26 | 0.50 | 99.41 | 81.64 | 66.67 |
| iPhone 12 Pro | 28 | 0.52 | 99.22 | 76.56 | 62.30 |

As shown in Table IV, the Google Pixel phones generate the highest accuracies for all three classification tasks. The iPhones generate slightly better results than Samsung phones. Samsung S8+ generated the worst accuracies. We notice the videos of Samsung S8+ suffer from missing frames potentially due to internal processing issues. We observe that $\eta_{cap}$ has the strongest correlation where lower $\eta_{cap}$ provides the adversary with less information and consequently lower accuracies. We also notice that there exists a trend of newer camera modules having lower $T_r$, i.e., higher rolling shutter frequency, and thus lower $\eta_{cap}$. The question of this trend being usable as a mitigation technique is further analyzed in section VII. All the phones we tested achieved at least 99.22%, 76.56%, and 61.07% accuracies on gender, speaker, and digit recognition respectively. This suggests that the adversary is able to perform successful side channel attacks with high $\mathbf{Adv}_A$ (Section III) on a large portion of phones available on the market at the time of writing.

**Multi-device Scalability.** To investigate the feasibility of cross-device attacks, we conducted four multi-device studies: (1) with the most recent phones from the three phone families (2) the four models of the Pixel family (3) the three models of the Samsung family (4) the three models of the iPhone family. As shown in Tables X, XI, XII, XIII, most cross-device cases show advantages over a random-guess adversary, demonstrating the existence of common information across different phone models. It is worth noting that when the classification model is trained on Pixel 5 and iPhone 12 Pro and tested on Samsung S20+, the accuracies for gender and digit recognition (highlighted in green in Table X) are higher than training on S20+ itself. Similarities in recovered signals across different models are determined by various sources, such as similar image sensors, rolling shutter frequencies, and image signal processing units (ISPs). For example, Samsung S7 and Pixel 1 have the same rolling shutter frequency and very similar ISP and processor. In contrast to the IMX260 sensor used in Samsung S7, the IMX378 sensor used in Pixel 1 does not support OIS [15]. The results suggest our side channel has the potential to be generalized for unseen devices, especially when the adversary trains with data from smartphone models with similar camera systems. The bolded numbers in Tables X, XI, XII, XIII indicate there is often no accuracy loss in testing on a specific phone when data from other phones are added to the training set. When training on all three or four phones and testing on a specific phone, the accuracies are almost ubiquitously better than or similar to training on that phone alone. This suggests our classification model is capable of representing data distribution from multiple phone models with minimal to no information loss.

## VII. DEFENSE

This section analyzes immediate countermeasures that may be carried out by users and more informed protections for manufacturers that aim to secure future camera devices.

### A. User-based Countermeasures

**Lower-quality Cameras.** Users can use lower-quality cameras to limit information embedded in videos by reducing video resolution and frame rate. However, these measures cannot degrade eavesdropping performance without significantly sacrificing overall video quality. Figure 6 (a) shows that reducing video resolution from $1080 \times 1920$ to $480 \times 640$ reduces the signal amplitude by about 60%. However, Figure 6 (b) and Table I show that when the volume decreases by 10 dB, the signal amplitude decreases by about 75% which only reduced digit classification accuracy from 79.69% to 76.95%.

**Phones Away From Speakers.** A straightforward yet effective approach for privacy-aware users is to place phones away from electronic speakers. As shown in Table VII, removing phones from the same surface as the speaker immediately reduces attack performance.

**Adding Dampening Materials.** Another possible method is to add vibration-isolation dampening materials between the phone and the surface in the hope to lower $k_0$ in Equation 2. Using the evaluation baseline setup and Pixel 2, we tested specialized vibration reduction mats made of visco-elastic polyurethane [17] with varying degrees of hardness. Three mats were used with common type OO durometers of 30, 50, and 70 [6]. Our tests show the three materials produced similar effects in mitigating our attack (Table VII). A classification evaluation shows adding such dampening materials reduced digit classification accuracies by 14.33% (Table V).

### B. Camera Design Improvement

Fundamentally, the side channel arises because of movable lenses that modulate smartphone motion onto video streams and rolling shutters that increase the available sample rate of adversarial signal recovery. We thus investigate the possible ways to mitigate these two sub-problems from the perspective of future camera designs.

*1) Rolling Shutter Mitigation:* Besides a plain approach of replacing rolling shutters with global shutters, we identify two methods to tackle the problem by increasing rolling shutter frequencies or adding randomization.

**Higher Rolling Shutter Frequency.** As mentioned in Section VI-D, we observed a trend of higher rolling shutter frequencies in newer camera sensors. We believe this trend shows camera designers' intention to approximate global shutters, which also led to lower attack performance as a byproduct. It is thus worth investigating the effectiveness of utilizing this trend as a defense. Basically, higher rolling shutter frequencies reduce the amount of intra-frame motion signals captured by adversaries (Section IV-B). We generated model-based predictions[2] of the side channel adversary's success with increasing rolling shutter frequencies and used the evaluation samples of Pixel 2 as the baseline.

Table VIII shows the tested $\eta_{cap}$, the required rolling shutter frequency, and the classification accuracies. The result suggests that further increasing the sample rate does reduce classification accuracies, but the adversary still has a large advantage over random-guess adversaries even if they can only recover 0.1% of the signals at 32,400 kHz. Furthermore, the accuracy decay sees an asymptotic trend, suggesting a potential lower bound of the accuracies even when the sample rate approaches infinite. We believe this lower bound is posed by the inter-frame information retained. In other words, adversaries may recover a large amount of information even from a global shutter camera just by measuring variations between frames.

**Random-coded Rolling Shutter.** If higher rolling shutter frequencies cannot be achieved, another method is to scramble the intra-frame signals by randomly mapping $s(n\delta)$ to $a(i)$ in Equation 7. Simply put, we can potentially randomize the order of each row's exposure and readout within each frame. This method only has a small impact on video quality because it only affects rolling shutter patterns in the videos which are already considered as distortions. Our simulation shows random-coded rolling shutter is able to produce defense effectiveness as good as increasing the rolling shutter frequency from 34 to about 100 kHz for Pixel 2. We conjecture this is because the intra-frame motion signals are only scrambled instead of completely removed and our classification model is able to utilize statistical information (e.g., max/min/mean) of the scrambled signals.

To implement random-coded rolling shutters, the address generator (Figure 2) needs to output randomly ordered instead of sequential addresses. Existing research shows manufacturers can already make the address generator output designated control sequences by changing camera firmware [36], [61]. The remaining cost of implementation is for adding a random number generator (RNG) that communicates with the address generator. In fact, imaging sensors themselves are a good source of entropy and have been already used in research and industry for generating true random numbers [3], [45], [80].

*2) Lens Movement Mitigation:* Our experiments show that addressing problems caused by rolling shutters alone cannot eliminate the threats due to the upper bound of protection effectiveness posed by the inter-frame motion information that still resides in the videos. It appears that the main cause of this side channel is the design flaw in existing smartphone camera sensors that leaves the lens dangling and free to move in the lens suspension system. Below, we propose two possible methodologies in an attempt to mitigate this.

TABLE V: Effectiveness of Single and Combined Defenses

| Defense | Gender (%) | Speaker (%) | Digit (%) |
|---|---|---|---|
| None (Baseline) | 99.87 | 91.02 | 79.69 |
| ① Rubber Mat Dampening | 98.64 | 80.11 | 65.36 |
| ② Higher RS Freq. (648 kHz) | 93.29 | 62.89 | 48.89 |
| ③ Random-coded RS | 98.18 | 76.56 | 60.22 |
| ①+② | 75.65 | 43.88 | 33.14 |
| ①+③ | 72.66 | 46.03 | 37.63 |
| ④ Tough Spring/Lens Locking | 65.23 | 16.73 | 16.67 |
| ②+④ | 53.91 | 8.66 | 16.73 |
| ③+④ | 54.36 | 8.46 | 13.93 |

**Tougher Springs.** Our signal path modeling reveals that increasing the elastic force of the lens suspension springs ($c_l$ in Equation 2) makes it more difficult for sound waves to vibrate the lens. There are several possible modifications designers can make to achieve this as suggested by the model of smartphone camera lens voice coil motor (VCM) systems [29]:

$$\begin{cases} S = \frac{R}{V^2} \cdot \left( \frac{F_e - f_{\text{fric}} - xc_l - mg}{m} \right)^2 \\ F_e = Nil_w B_g = N \frac{VA}{\rho L} l_w B_g \end{cases}$$

$S$ is the VCM's sensitivity that designers want to optimize; $F_e$ is the electromagnetic actuation force; $x$ is the lens displacement. To keep $S$ the same so that users do not experience degradation in camera functionality and usability, we identify the following straightforward ways to compensate for the impact of higher $c_l$ along with their costs. (1) Increase the number of coil windings $N$, the coil length $l_w$, or coil area $A$. This will increase the size of the camera modules. (2) Increase the magnetic flux density $B_g$ by using better permanent magnet materials. This will add to the budget. Other parameters such as coil voltage $V$ and resistance $R$ may also be adjusted but can lead to higher camera power consumption. While different camera products are subjected to specific manufacturing constraints, we believe our analysis above provides a starting point that designers can consider.

**Lens Locking.** We envision the ultimate solution to the lens movement problem is to have a locking mechanism that completely prevents lenses from moving when they are not supposed to. Such a mechanism may be achieved by adding controllable pillars around the lens. The pillars contract when OIS and AF are enabled to make space for lens movement and expand to fix the lens in place otherwise.

**Simulation of Effectiveness.** To demonstrate the potential of these solutions, we simulated tougher springs and lens locking by using an external magnet to prevent the lens from moving in the same way as Section IV-A. The decreasing attack accuracies are shown in Table V. The remaining non-random classification accuracies are likely due to a combination of (1) the residual lens movements that the magnet cannot completely remove, and (2) the tiny movements of the smartphone body. Finally, combining multiple methods of defense can further bring attack performance down to the random-guess range as shown in Table V.

## VIII. Discussion & Future Work

**Limitations.** The main limitation of this side channel is its dependence on a suitable mechanical path from the sound source to the smartphone that allows strong structure-borne sound wave propagation. It is worth noting that the requirement of having structure-borne propagation paths itself does not pose more limitations than air-borne propagation since the two types of propagation almost always coexist except in zero-gravity environments [30]. Air-borne propagation does not enable this side channel attack with present-day smartphone cameras because the kinetic energy it transfers is insufficient to vibrate the phone to such a level that discernible pixel displacements can be generated. From Eq. 2, this means $E_p$ is too small due to a small $k_0$ for air-borne propagation while structure-borne propagation often has much larger $k_0$. On the other hand, the major limitations posed by structure-borne propagation are two-fold.

First, the $k_0$ of structures is more complex, less homogeneous, and thus harder to predict and model than that of air. For example, different table materials produce different signal quality as shown in Table I. Second, the $k_0$ of structures can also be smaller than what is needed to cause discernible signals in videos in certain scenarios. For example, we found speech from human speakers is not able to produce discernible signals on phones held in their hands or placed in front of them on a table. We believe this is because human body is generally a bad mechanical wave transfer medium with its soft tissues reducing $k_0$.

**Future Directions.** We believe the high classification accuracies obtained in our evaluation and the related work using motion sensors [22] suggest this optical-acoustic side channel can support more diverse malicious applications by incorporating speech reconstruction functionality in the signal processing pipeline. In addition, future research can also consider combining signals from multiple cameras to combat the problem of lower $\eta_{cap}$. To achieve this, an advanced adversary needs to address the challenge of building a recognition model that can tolerate the randomness in the exposure time difference between different cameras.

## IX. Related Work

**Sound Recovery From Vibrating Objects In Videos.** The concept of recovering sound by analyzing vibrating objects in video frames was first introduced by Akutsu et al. [19] in 2013 where they used high-speed cameras (over 6,000 fps) to record the movements of a speaker's face and neck. Davis et al. [32] found it is possible to recover speech by aiming a specialized high frame-rate camera at lightweight objects (e.g., plastic bags) vibrated by sound waves. Follow-up research on this topic mainly focused on improving the efficiency of sound recovery based on Davis's technique using specialized high frame-rate cameras [78], [79], [81]. Some works also discussed the possibility of utilizing the rolling shutter effect to emulate higher frame rates with common cameras, but the discussions remain proof-of-concept in lab settings as it requires a high-end camera on a tripod to focus precisely on the lightweight objects at a very close distance [32], [33], [58], [76]. In comparison, our work exploits the rolling shutter artifacts caused by the movement of smartphone camera lenses that are intrinsic to existing smartphone camera hardware itself. This feature allows our optical-acoustic side channel to work without any vibrating object in the camera's field of view and enabled us to evaluate a wide range of possible sound recovery scenarios including when the speaker and camera are in two different rooms (Section VI-B). Furthermore, previous works' recovered signal amplitude is proportional to the lens focal length due to their need of objects in the video frames, which poses the major limitation of requiring short camera-object distance or expensive optics [32]. In comparison, our work addresses this limitation by exploiting the movable lens structure on smartphone cameras as a signal amplifier under structure-borne sound.

**Smartphone Motion Sensor Side Channels.** In 2014, Gyrophone [46] first proposed the idea of using gyroscopes on smartphones for acoustic eavesdropping. They investigated a structure-borne attack scenario where the smartphone and electronic speaker are on a shared table surface. Following works such as AccelEve [22], Spearphone [21], and [27] proposed a structure-borne threat model of eavesdropping audio played by the smartphone's built-in electronic speakers with accelerometers on the same phone, which is similar to our same-phone scenario evaluated in Section VI-A.

Compared to motion sensor side channels, the optical-acoustic side channel proposed in this work opens up a new modality of smartphone acoustic eavesdropping since cameras create an orthogonal space of threat models in cases where motion sensor data is not available or add to the total information extracted when it coexists with motion sensors. Camera side channels provide a high bandwidth while motion sensors often have better sensitivity to vibrations. In addition to the shared surface-coupling and phone body-coupling scenarios, our paper further investigates new scenarios where smartphones are on different surfaces than the speakers such as on a different desk, in a shirt pocket, in a bag, or even in a different room. It is worth pointing out that comparisons between these motion sensor side channels' results and our results may not provide meaningful insights due to the large differences in their threat models, algorithms, evaluation setups, etc.

**Physical Acoustic Eavesdropping.** Researchers also exploited other physical mechanisms for acoustic eavesdropping. We refer the readers to the SoK paper by Walker et al. [71] for a relatively comprehensive review. Lamphone [48] and the little seal bug [49] use telescopes and optical sensors to sense the optical changes caused by sound-induced object vibrations remotely. Glowworm [47] finds that the LED light intensity of electronic speakers leaks acoustic information and uses telescopes and photodiodes to eavesdrop on it. Compared to these works, our work does not require specialized devices and light but uses smartphones in private spaces for eavesdropping. LidarPhone [55] inherits the well-studied concept of sound laser vibrometry but uses malware to exploit the lidar sensors on robot vacuum cleaners for eavesdropping. Hard drive of hearing [43] discovers that the read/write head of hard drives can be turned into unintentional microphones for eavesdropping when the head is vibrated by loud sounds.

**Camera-based Attacks.** Poltergeist [40] by Ji et al. studied the robustness problem of the camera OIS from an almost complementary perspective to our work. They discovered that adversaries can generate intentional ultrasounds to change the gyroscope readings of OIS in similar ways as explored in [59], [67] and thus cause controlled motion blurs in the camera videos to attack computer vision-based autonomous vehicles. They See Me Rollin' by Köhler et al. [42] studied laser-based optical injection attacks against CMOS cameras in autonomous vehicles. They exploited the rolling shutter mechanism of CMOS cameras to inject row-wise fine-grained disruption patterns into camera videos that could hide up to 75% of objects perceived by state-of-the-art computer vision object detectors. While their work studies how rolling shutters can cause robustness and security problems to downstream processing units when subjected to active optical injections, our work investigates how to passively recover ambient acoustic information from rolling shutter cameras vibrated by sound.

## X. Conclusion

In this work, we investigated the threat of point-of-view acoustic signal eavesdropping from smartphone cameras' image streams. By investigating the rolling shutter and movable lens architectures, we examined the side channel's feasibility and limits under different setups and scenarios. Our analysis and experiments with 10 smartphones demonstrate how malicious parties with knowledge of camera hardware structure can extract fine-grained acoustic information from recorded videos, achieving digit, speaker, and gender recognition. We investigated user-based countermeasures and possible camera design improvements that address the problems caused by rolling shutters and movable lenses. Simulation results show that different strategies can be combined to reduce attack performance to the random-guess range.

## Acknowledgements

This work is supported by a gift from Analog Devices and a gift from Facebook. We thank our reviewers for their highly constructive comments and Erica Eaton for her help in iOS app development, data collection, and proofreading; We also thank Wenyuan Xu and Xiaoyu Ji for their insightful feedback.

# APPENDIX A
## CAMERA HARDWARE & RECORDING APP

**OIS, EIS, and AF.** Both OIS and EIS try to eliminate the image blurs caused by low-frequency human tremors (<20 Hz). OIS senses the human tremor with an internal gyroscope in the camera module and directly moves the optical modules, such as the lens, for compensation. Most tremor-caused motions result in approximately translational transformations of images, which can be compensated by just 2D movements of the lens that are parallel to the pixel array plane. The movable lenses in many OIS systems are thus designed to have a degree of freedom (DoF) of 2 [44], [57], [73]. EIS uses the gyroscope of the phone itself and tries to correct image blurs by operating in the software domain. The effects of leaving OIS and EIS on when recovering a chirp is as shown in Figure 9 (b) and (c). Similarly, AF introduces extra distortions when left on because the sound-induced lens motions can change the focus of the camera at which point AF tries to intervene.

TABLE VI: Objects On The Glass Desk

| Objects | Volume (dB) | Avg. SNR | Avg. STOI |
|---|---|---|---|
| Speaker | 85 | 4 | 0.24 |
|  | 65 | 1.7 | 0.18 |
| Speaker+Laptop | 85 | 3.5 | 0.24 |
|  | 65 | 1.7 | 0.2 |
| Speaker+Laptop +Monitor | 85 | 3.8 | 0.25 |
|  | 65 | 1.6 | 0.19 |
| Speaker+Laptop +Monitor+Printer | 85 | 2.6 | 0.21 |
|  | 65 | 1.6 | 0.2 |

**Camera Control Parameters.** Summarized as follows.

(1) Disable auto-exposure and reduce the exposure time (Section IV-B). Practically, the adversary cannot arbitrarily reduce the exposure time since that also reduces the amount of light captured. Considering this trade-off and our observation that higher frequencies are often limited by the mechanical transfer function instead of the exposure time, we use a 1 ms exposure time that generally strikes the best balance. Figure 9 (d) shows the recovered chirp under 10 ms as a comparison.

(2) Disable optical and electronic image stabilization (OIS and EIS) and auto-focus (AF). When OIS, EIS, and AF are left on, their control systems will cause extra distortions in the images that do not reflect the true sound-induced lens motion (see Appendix A for the details), as shown in Figure 9 (b), (c), (e) respectively. A considerate adversary should thus disable these functionalities when performing eavesdropping.

(3) Minimize video codec compression. The adversary should avoid distortions introduced by the video codec by disabling inter-frame compression, increasing the average bit rate, etc. In comparison to Figure 9 (e), Figure 9 (f) shows the chirp recovered with the smartphone's stock camera app where all parameters above including the codec are not optimized. This represents what a naive adversary can recover without the knowledge developed in Section IV-A and IV-B.

(4) Increase pixel resolution. Higher pixel resolution enables the recovery of smaller-amplitude lens motions according to Equation 4. From another perspective, pixel displacement amplitude increases with the resolution given a constant-amplitude lens motion. For example, Figure 6 shows the measured amplitude of a 200 Hz signal under common video resolutions where the signal amplitude sees an almost linear increase with the number of columns/rows of the frames.

(5) Choose appropriate frame rates. When $T_r$ and $T_e$ remain constant, increasing the frame rate increases the amount of captured signal (as shown in Section IV-B). But when camera sensors try to output unusually high frame rates (e.g., in the slow motion mode), they tend to reduce $T_r$ to accommodate the exposure of more frames. We found this can lead to the same or even smaller amount of captured signal at the cost of wasting extra computation resources. We use 30 and 60 fps for Android and iOS devices respectively.

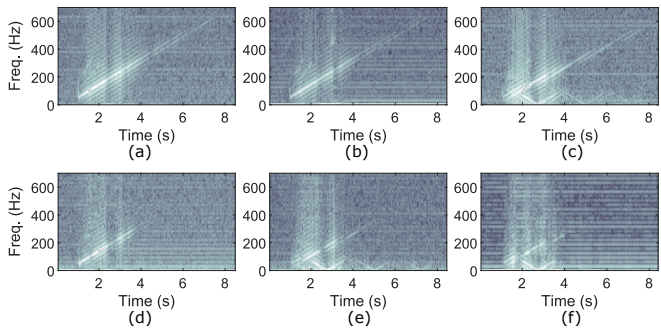

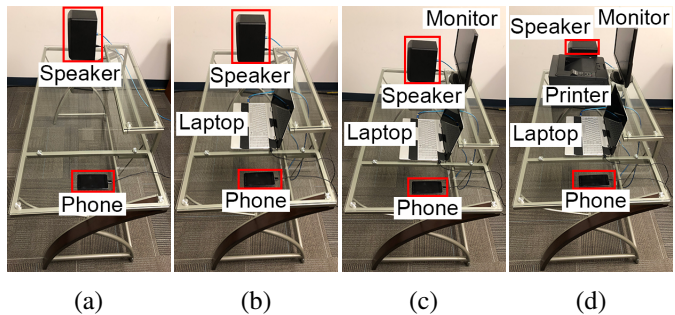

Fig. 9: The recovered chirp signals (50-650 Hz in 7s) with different camera control parameters and a 30 fps frame rate. (a) Optimized parameters and 1 ms exposure time. (b) OIS is left on. (c) EIS is left on. (d) 10 ms exposure time. (e) OIS, EIS, AF are left on with 10 ms exposure time. (f) Recovered with the phone stock camera app without any optimization.

Fig. 12: Setups for evaluating additional objects sequentially added onto the desk between the phone and speaker. (a) Only the phone and speaker are on the desk. (b) A laptop is added. (c) A monitor is added. (d) A printer is added.

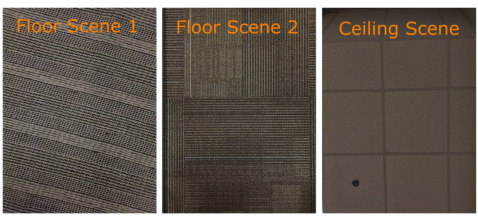

Fig. 10: The three scenes evaluated.

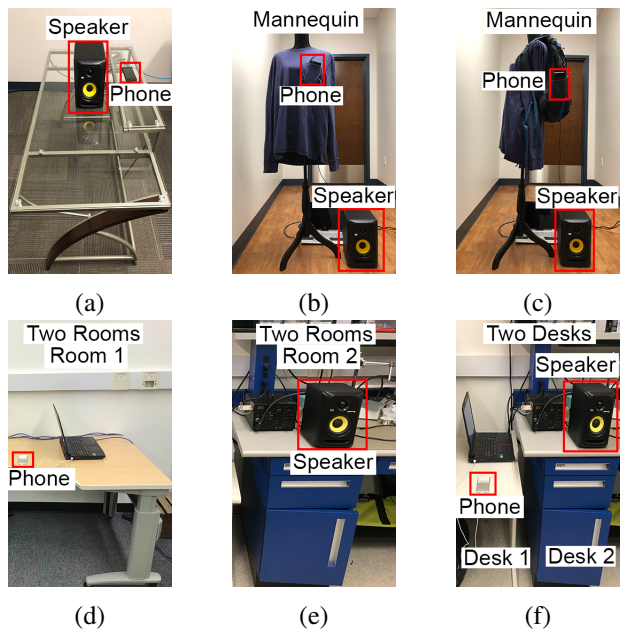

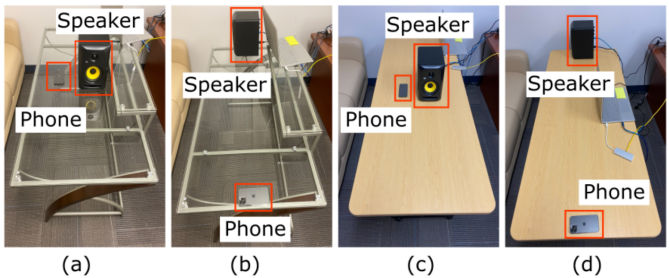

Fig. 11: Setups of glass and wooden desks with the camera facing the ceiling. From the left. (a) 10 cm phone-speaker distance (b) 110 cm phone-speaker distance (c) 10 cm phone-speaker distance (d) 130 cm phone-speaker distance.

Fig. 13: Setups of different-surface scenarios. (a) The phone is on the monitor stand. (b) The phone is in the pocket on a mannequin. (c) The phone is in the pocket of a backpack. (d, e) The phone and speaker are in two different rooms. (f) The phone and speaker are on two different desks.

## APPENDIX B
## SIDE CHANNEL PERFORMANCE ANALYSIS

To help readers understand the different adversary performances observed in various scenarios (Section VI), we provide further analyses on how the propagation surface materials and structures affect the performance of the side channel.

**Energy Transformation.** Since the law of conservation of energy applies to the vibration energy $E_s$, any change in the experimental setup that leads to frictional losses will degrade the performance of the side channel. For instance, in the case of different surface experiments, the contact point between the objects, such as backpack and mannequin, the mannequin's foot, and the floor, and desk and floor, will cause frictional losses from $E_s$ which would otherwise be transformed to kinetic energy that vibrates the target smartphone. We believe this is why we observed worse adversary performance in different-surface scenarios than shared-surface scenarios.

**Wooden Desk Phone-speaker Distance.** Higher accuracies were observed with larger phone-speaker distances on the wooden desk (Section VI-B). The wooden desk has a metal brace screwed underneath the desk's structure. We believe that this increases the effective thickness of the desk, which increases the second moment of area of the desk by a cubic factor [28]. The second moment of area is proportional to the stiffness of the desk, which significantly decreases the vibration-borne deflections. In addition, the material's elastic modulus is inversely proportional to the deflection of the structure and the elastic modulus of the metal brace is significantly larger than wood, resulting in a further decrease in deflection. When the phone is placed on one end of the desk and the speaker on the other end, $E_s$ propagates through the section with a larger effective thickness to the edge of the desk that does not have this additional brace. This results in lower stiffness at the edges, which leads to larger deflection and thus larger phone camera lens movements.

TABLE VII: Vibration Reduction Mats

| Durometer | Avg. SNR | Avg. STOI |
|---|---|---|
| 70 (OO) | 12.43 | 0.46 |
| 50 (OO) | 12.00 | 0.42 |
| 30 (OO) | 11.47 | 0.44 |

TABLE VIII: Recognition Accuracy With Different $\eta_{cap}$

| $\eta_{cap}$ (%) | Sample Rate (kHz) | Gender (%) | Speaker (%) | Digit (%) |
|---|---|---|---|---|
| 95 | 34 | 99.87 | 91.02 | 79.69 |
| 50 | 65 | 99.54 | 80.86 | 59.77 |
| 10 | 324 | 95.51 | 68.55 | 50.59 |
| 5 | 648 | 93.29 | 62.89 | 48.89 |
| 1 | 3,240 | 86.78 | 48.37 | 41.21 |
| 0.5 | 6,480 | 86.85 | 43.88 | 38.87 |
| 0.1 | 32,400 | 83.20 | 42.25 | 38.54 |

**Floor Propagation.** When $E_s$ is applied to more complex structures, such as wooden or concrete floors, the effective stiffness $K$ of the structures is often much larger than glass or wood used in desks, resulting in smaller surface deflection. This decrease in structural deflection results in vibrations with smaller amplitudes and smaller SNR, which degrade the side channel performance (Section VI-C).

TABLE IX: Information of Different Smartphone Models Tested

| Device | Rear Cam. Module | $1/T_r$ (kHz) | Max Chirp Freq. (Hz) | Meas. $\eta_{cap}$(%) (Calc. $\eta_{cap}$)(%) | OS Version | Device Released | Processor SoC | ISP on Processor SoC |
|---|---|---|---|---|---|---|---|---|
| Pixel 1 | IMX378 | 45 | 600 | 72 (72) | Android 10 | Oct. 2016 | Qualcomm 821 | Spectra Dual ISPs |
| Pixel 2 | IMX362 | 34 | 600 | 96 (95) | Android 11 | Oct. 2017 | Qualcomm 835 | Spectra 180 |
| Pixel 3 | IMX363 | 34 | 600 | 97 (95) | Android 9 | Oct. 2018 | Qualcomm 845 | Spectra 280 |
| Pixel 5 | IMX586 | 58 | 600 | 57 (56) | Android 9 | Sep. 2020 | Qualcomm 765G | Spectra 355 |
| Samsung S7 | IMX260 | 45 | 600 | 73 (72) | Android 7 | Mar. 2016 | Qualcomm 820 | Dual ISPs |
| Samsung S8+ | IMX333 | 45 | 600 | 73 (72) | Android 8 | Apr. 2017 | Qualcomm 835 | Spectra 180 |
| Samsung S20+ | IMX555 | 58 | 650 | 56 (56) | Android 11 | Mar. 2020 | Qualcomm 865 | Spectra 480 |
| iPhone 7 | Unknown | 92 | 600 | 72 (70) | iOS 15.3 | Sep. 2016 | A10 Fusion chip | Unknown |
| iPhone 8+ | Unknown | 92 | 650 | 72 (70) | iOS 15.3 | Sep. 2017 | A11 Bionic chip | Unknown |
| iPhone 12 Pro | Unknown | 160 | 600 | 42 (40) | iOS 15.1 | Oct. 2020 | A14 Bionic chip | Unknown |

TABLE X: Multi-device Training and Testing: Cross-band

| Trained \ Tested | Gender (%) | | | Speaker (%) | | | Digit (%) | | |
|---|---|---|---|---|---|---|---|---|---|
| | Px5 | S20+ | 12 Pro | Px5 | S20+ | 12 Pro | Px5 | S20+ | 12 Pro |
| All Three Phones | **99.48** | **99.89** | **99.67** | **85.42** | **83.92** | **83.59** | 69.66 | **70.31** | 71.74 |
| Pixel 5, iPhone 12 Pro | 99.54 | 99.87 | 99.54 | 84.24 | 79.04 | 81.64 | 61.98 | 62.83 | 61 |
| Pixel 5 | 99.48 | 59.9 | 54.3 | 84.51 | 12.43 | 7.49 | 70.25 | 24.67 | 18.42 |
| Samsung S20+ | 62.24 | 99.8 | 74.15 | 5.21 | 83.92 | 9.51 | 23.44 | 61.07 | 17.45 |
| iPhone 12 Pro | 66.8 | 66.6 | 99.22 | 8.72 | 11.59 | 76.56 | 18.16 | 19.79 | 62.3 |

TABLE XI: Multi-device Training and Testing: Google Pixel Phones

| Trained \ Tested | Gender (%) | | | | Speaker (%) | | | | Digit (%) | | | |
|---|---|---|---|---|---|---|---|---|---|---|---|---|
| | Px1 | Px2 | Px3 | Px5 | Px1 | Px2 | Px3 | Px5 | Px1 | Px2 | Px3 | Px5 |
| Px1, Px2, Px3, Px5 | 99.02 | 99.8 | **99.67** | 99.22 | **85.16** | **91.6** | 89.32 | 83.27 | **75.39** | **81.84** | 76.76 | 70.18 |
| Px1, Px2, Px5 | 98.63 | 99.61 | 98.89 | 99.35 | 84.64 | 90.95 | 67.38 | 83.4 | 73.83 | 81.12 | 47.01 | 68.62 |
| Px2, Px5 | 57.81 | 99.41 | 98.83 | 99.28 | 13.35 | 91.8 | 65.95 | 83.53 | 17.38 | 81.9 | 43.03 | 69.73 |
| Px1 | 99.61 | 59.57 | 67.64 | 63.67 | 81.84 | 5.6 | 5.21 | 6.25 | 69.53 | 28.65 | 24.48 | 11.91 |
| Px2 | 56.18 | 99.87 | 99.41 | 93.36 | 7.1 | 91.02 | 55.79 | 22.27 | 27.93 | 79.69 | 41.34 | 20.38 |
| Px3 | 59.05 | 97.01 | 99.67 | 95.12 | 8.66 | 52.02 | 91.28 | 22.59 | 24.09 | 59.57 | 80.66 | 23.76 |
| Px5 | 62.7 | 67.51 | 81.45 | 99.48 | 10.61 | 19.47 | 21.35 | 84.51 | 13.67 | 13.41 | 11.72 | 70.25 |

TABLE XII: Multi-device Training and Testing: Samsung Galaxy Phones

| Trained \ Tested | Gender (%) | | | Speaker (%) | | | Digit (%) | | |
|---|---|---|---|---|---|---|---|---|---|
| | S7 | S8+ | S20+ | S7 | S8+ | S20+ | S7 | S8+ | S20+ |
| S7, S8+, S20+ | 99.15 | 98.96 | 99.28 | **88.74** | 75.2 | 81.58 | **75.78** | **58.66** | **66.54** |
| S7, S20+ | 99.15 | 94.4 | 99.61 | 86.33 | 39.71 | 81.71 | 75.46 | 38.09 | 64.97 |
| S7 | 99.54 | 93.95 | 57.88 | 82.94 | 32.94 | 5.21 | 66.08 | 31.38 | 10.87 |
| S8+ | 94.34 | 99.61 | 27.15 | 47.66 | 79.3 | 5.01 | 43.03 | 57.29 | 20.44 |
| S20+ | 50.26 | 49.8 | 99.8 | 5.4 | 4.3 | 83.92 | 10.87 | 12.24 | 61.07 |

TABLE XIII: Multi-device Training and Testing: Apple iPhones

| Trained \ Tested | Gender (%) | | | Speaker (%) | | | Digit (%) | | |
|---|---|---|---|---|---|---|---|---|---|
| | iP7 | iP8+ | iP12 Pro | iP7 | iP8+ | iP12 Pro | iP7 | iP8+ | iP12 Pro |
| iP7, iP8+, iP12 Pro | 99.41 | 99.35 | 98.89 | 83.33 | **83.92** | 81.51 | 68.82 | **72.33** | **66.34** |
| iP7, iP12 Pro | 99.35 | 80.73 | **99.28** | 83.2 | 19.34 | **79.95** | **69.7** | 18.62 | 65.36 |
| iP7 | 99.87 | 39.71 | 49.09 | 85.09 | 5.73 | 5.14 | 65.23 | 13.74 | 18.75 |
| iP8+ | 49.93 | 99.41 | 79.36 | 6.9 | 81.64 | 14.78 | 12.24 | 66.67 | 12.43 |
| iP12 Pro | 50.13 | 53.26 | 99.22 | 5.73 | 10.61 | 76.56 | 15.1 | 14.32 | 62.3 |

# Supplementary Materials

## MOVING MEAN FILTER

The transfer function of the moving mean filter revealed by Equation 7 is:

$$H(\omega) = (1/L)\left(1 - e^{-j\omega L}\right) / \left(1 - e^{-j\omega}\right)$$

where $L = T_e/\delta$ stands for the number of discrete points being averaged. The moving mean filter's frequency response is thus determined by exposure time $T_e$, as shown in Figure 14.

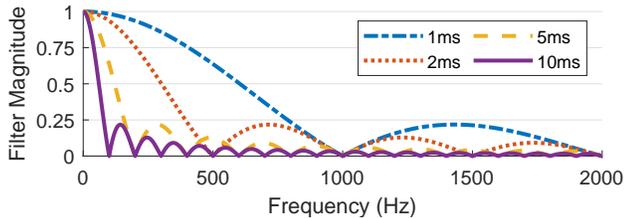

Fig. 14: The signal sampling process (Section IV-B) from rolling shutter patterns acts as a moving mean filter with the exposure time as the key variable. Shorter exposure times support larger signal bandwidth.

## CAMERA RECORDING APPS

**Android Data Collection App.** The Android data collection app is adapted from the app introduced in [39]. It uses the Camera2 and MediaCodec APIs provided by Android for controlling camera and video codec parameters respectively. Besides the camera controls specified in Section IV-D, we use a 1080 × 1920 resolution and a 30 fps frame rate for all evaluations. The video codec is h264 with all frames set to be key frames and the average bit rate set to provide a bit per pixel value of 1. It is worth noting that although Android exposes APIs for frame rates higher than 30 fps in normal (non-slow motion) recording mode, some smartphone manufacturers including Google and Samsung currently do not support 3rd party apps to do it.

**iOS Data Collection App.** The iOS data collection app is modified from the AVCam app provided by Apple Developer documentation [18]. The app uses the AVFoundation framework to record video from the iPhone's camera. In addition to the camera controls detailed in Section IV-D, we use 1080×1920 resolution and a frame rate of 60 fps. This frame rate was chosen because it doubled the value of $\eta_{cap}$, providing the value for $\eta_{cap}$ shown in Table IV without entering the slow motion mode. We use the JPEG video codec with video quality set to 1.0, which prevents frame compression. With the JPEG video codec, each frame is treated as a key frame. Similar to the Android app, the average bit rate is set to provide a bit per pixel value of 1.

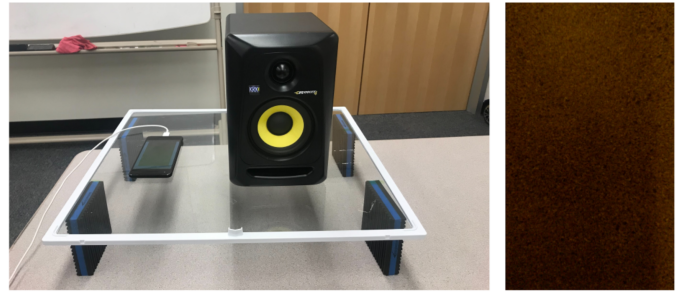

Fig. 15: (Left) The feasibility test (Section IV-D) setup. (Right) The scene of the smartphone's rear camera.

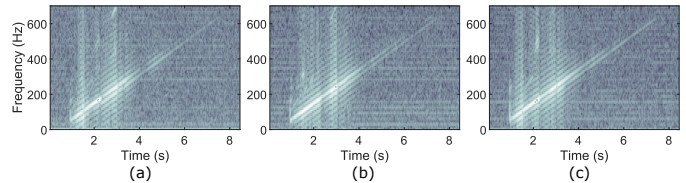

Fig. 16: The chirp signal recovered with different orientations (Section IV-D) of the phone with respect to the sound source (a) 0 degrees. (b) 45 degrees. (c) 90 degrees. We observe no significant variations for different orientation.

## CLASSIFICATION MODEL

Previous research on acoustic side-channel that use smartphone motion sensors for speech eavesdropping utilized a variety of machine learning models, such as Hidden Markov Models (HMM) [46], Dynamic Time Wrapping (DTW) [46], XGBoost [27], and Random Forest [21], [27], to recognize and classify speech. Such recognition models often utilize features including mel-frequency cepstral coefficients (MFCCs) or speech spectrograms. Different from these previous works, we designed a HuBERT-based waveform input deep learning classification model (Section V-B). Figure 17 displays the architectural design of the HuBERT-based classifier model developed for our optical-acoustic side-channel. Here we provide our reasoning for this design.

**Transfer Learning.** HuBERT is a machine learning model that implements the concept of transfer learning. Essentially, the acoustic signals we extract are the embeddings of the original speech signals in the lens motion domain. Although building classification models in such a new signal domain often requires a significant amount of data to sufficiently sample the target domain, the adversary can overcome this challenge by utilizing transfer learning. Simply put, transfer learning allows the model to leverage the knowledge gained in the source domain (speech audio) and apply it to the tasks in the target domain (lens motion). Specifically, the original

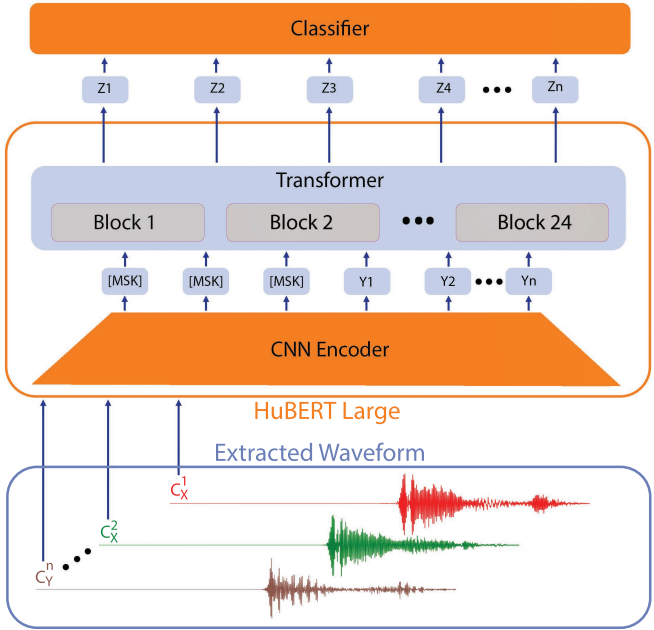

Fig. 17: HuBERT-based classifier model architecture (Section V-B). Three major components of this model are the CNN decoder, transformer, and classifier. The input channels of the CNN decoder are modified to allow for all extracted signals to be used by the model for inference.

HuBERT model was pre-trained on 60,000 hours of speech derived from the LibriVox project and formalized in Librilight [41], as well as 960 hours of speech from Librispeech [50]. This large amount of incorporated knowledge enables the model to rapidly adapt to the lens motion distribution.

**Waveform Input.** HuBERT uses audio waveform as inputs. Compared to other popular types of speech recognition inputs, such as MFCCs and signal spectrogram, waveform inputs preserve the largest amount of usable information without being transformed under any potentially biased hypothesis regarding the data distribution that can cause loss of valuable information. Audio waveform inputs have also been demonstrated in speech recognition research to achieve higher performance in a variety of scenarios [23], [37], [38], [56].

**State-of-the-art.** HuBERT is a waveform transformer deep learning model with embedding networks [56] and an attention mechanism [24] developed for speech recognition [23], [37], [38]. Currently, wave transformer models have produced the state-of-the-art performance in a variety of speech recognition tasks. Waveform transformer models perform better than convolutional and recurrent neural networks in speech recognition by leveraging self-supervised and supervised pre-training on large speech corpora to learn various phoneme permutations corresponding to complex human speech. In our experimentation, we observed better performance of HuBERT on our recognition tasks compared to a variety of other network models such as ResNet and EfficientNet [35]. We further experimented with Hubert's base, large, and x-large models and found that optimal results for our datasets can be achieved with HuBERT large.

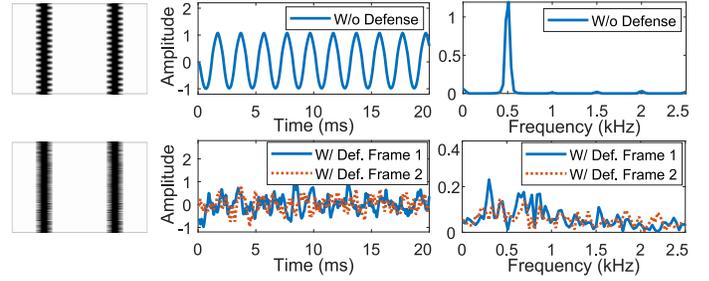

Fig. 18: The simulated result of random-coded rolling shutter (Section VII-B1) as a defense. (Upper) Without the defense, a 500 Hz signal can be easily extracted. (Lower) With the defense, the 500 Hz signal gets randomly projected to different indiscernible signals in different frames.

ROLLING SHUTTER SAMPLE RATE SIMULATION

Algorithm 1 shows the proposed algorithm for generating model-based predictions of sample data with increasing rolling shutter frequencies (Section VII-B1). It transforms the original waveform $W$ to $W'$ using the desired rolling shutter sampling rate $S_r$ to simulate the waveform generated by a rolling shutter with a different sampling rate. To transform $W$ to $W'$, we compute an upsampling factor $U$ by dividing the original captured signal percentage $\eta_{cap}$ by the desired capturing signal percentage $\eta_d$. Given that the phone used in this experiment is Pixel 2 its captured percentage or $\eta_{cap}$ can be calculated using equation 5 where the $M$ or number of rows in each frame is 1080 in the original data and the $f_v$ or frame rate is 30 fps. $\eta_d$ can be calculated using the equation below:

$$\eta_d = \frac{1080 \times 30}{S_r} \quad (8)$$

Using the sampling rate or $\frac{1}{T_r}$ is 34000Hz according to Table IV, the upsampling factor or $U$ can be computed using the equation below:

$$U = \frac{1080 \times 30}{34000\,\eta_d} \quad (9)$$

---

**Algorithm 1** Rolling Shutter Sample Rate Simulation

**Input:** $W$, $S_r$
**Output:** $W'$
1: Calculate $\eta_d$ using equation 8
2: Calculate $U$ using equation 9
3: Upsample $W$ to $U$ times the original length for all channels
4: Create a square wave with the duty cycle equal to $\eta_d$ and period equal to $\frac{1}{30}$ normalized between 1 and 0
5: Multiply the two signals
6: Concatenate the non-zero parts of the signal to create $W'$

DUROMETER OF DAMPENING MATERIALS

Durometer (Section VII-A) is metric that quantifies the hardness of a material [51] with a lower value representing a softer material and potentially a larger degree of dampening. Theoretically, softer materials (lower durometer values) often produce better vibration-isolation effects. However, an extremely soft material will not return its to original shape after deformation which is not suitable for vibration dampening. It is thus important to select the right materials. The three vibration reduction mats we tested have durometers of 30, 50, and 70 OO, which are the most common options for vibration reduction materials. These values approximately correspond to the hardness of gel shoe insole, rubber band, and pencil eraser respectively.